\documentclass{aa}
\usepackage{graphicx}
\usepackage[varg]{txfonts}
\usepackage{cases}
\usepackage{natbib}

\begin{document}
      \title{Grow-up of a Filament Channel by Intermittent Small-scale Magnetic Reconnection}
      
      \author{H. T. Li\inst{1,3}, X. Cheng\inst{1,2,3}, J. H. Guo\inst{1,3}, X. L. Yan\inst{4}, L. F. Wang\inst{1,3}, Z. Zhong\inst{1,3}, C. Li\inst{1,3}, M. D. Ding\inst{1,3}}
      
      \institute{School of Astronomy and Space Science, Nanjing University, Nanjing 210023, People$'$s Republic of China \\
      	\email{xincheng@nju.edu.cn} \\
 		\and
 		Max Planck Institute for Solar System Research, Gottingen, D-37077, Germany \\
 		\and
 		Key Laboratory of Modern Astronomy and Astrophysics (Nanjing University), Ministry of Education, Nanjing 210023, People$'$s Republic of China \\
 		\and
 		Yunnan Observatories, Chinese Academy of Sciences, Kunming 650216, People$'$s Republic of China \\
 	    }
    \titlerunning{Grow-up of a Filament Channel by Intermittent Small-scale Magnetic Reconnection}
 \authorrunning{Li et al.}
\date{Received; accepted}

\abstract
{Filament channel (FC), a plasma volume where the magnetic field is primarily aligned with the polarity inversion line, is believed to be the pre-eruptive configuration of coronal mass ejections. Nevertheless, evidence for how the FC is formed is still elusive.}
{In this paper, we present a detailed study on the build-up of a FC to understand its formation mechanism.}
{The New Vacuum Solar Telescope of Yunnan Observatories and Optical and Near-Infrared Solar Eruption Tracer of Nanjing University, as well as the AIA and HMI on board Solar Dynamics Observatory are used to study the grow-up process of the FC. Furthermore, we reconstruct the non-linear force-free field (NLFFF) of the active region using the regularized Biot-Savart laws (RBSL) and magnetofrictional method to reveal three-dimension (3D) magnetic field properties of the FC.}
{We find that partial filament materials are quickly transferred to longer magnetic field lines formed by small-scale magnetic reconnection, as evidenced by dot-like H$\alpha$/EUV brightenings and subsequent bidirectional outflow jets, as well as untwisting motions. The H$\alpha$/EUV bursts appear repeatedly at the same location and are closely associated with flux cancellation, which occurs between two small-scale opposite polarities and is driven by shearing and converging motions. The 3D NLFFF model reveals that the reconnection takes place in a hyperbolic flux tube that is located above the flux cancellation site and below the FC.}
{The FC is gradually built up toward a twisted flux rope via series of small-scale reconnection events that occur intermittently prior to the eruption. }

\keywords{Sun: activity -- Sun: filaments, prominences -- Sun: magnetic fields}

\maketitle

\section{Introduction}\label{introduction} 
Coronal mass ejections (CMEs) are large-scale explosive phenomena in the solar system and can give rise to severe space weather events \citep{2011LRSP....8....1C}. A promising method to forecast the occurence of CMEs is to monitor their pre-eruptive configurations including filaments, coronal cavities, sigmoids, and hot channels \citep{1995ApJ...443..818L, 1998GeoRL..25.2481H, 2010ApJ...719L.181W, 2012NatCo...3..747Z, 2013ApJ...769L..25C, 2014ApJ...789L..35C, 2019ApJ...887..130H}. In a sense, these pre-eruptive structures can be regarded as different manifestations of a common configuration, i.e., so-called filament channel (FC), at the different evolution stages and/or plasma environments as suggested by a recent review paper by \citet{2020SSRv..216..131P}.

Observationally, the FC corresponds to a plasma volume where the magnetic field is primarily aligned with the polarity inversion line (PIL). Usually, the magnetic field within the FC is highly sheared or twisted, which can be classified into two categories: sheared magnetic arcade (SMA; \citealp{1957ZA.....43...36K, 1994ApJ...420L..41A, 2000ApJ...539..954D}) and magnetic flux rope (MFR; \citealp{1974A&A....31..189K, 1996A&A...316..201W, 1998A&A...335..309A, 2006JGRA..11112103G, 2014ApJ...789L..35C, 2015ApJS..219...17Y}). The latter is usually believed to be more coherent and has a larger average twist (e.g., $\geq$ 1 turn). In many cases, the pre-eruptive FC manifests as a filament on the solar disk or prominence above the solar limb with the cool materials suspended at magnetic dips \citep{1974GAM....12.....T, 1994ApJ...420L..41A, 1998SoPh..182..107M, 2010SSRv..151..333M, 2014ApJ...784...50C, 2015ASSL..415..355M}. However, this is not always the case since some studies have shown that magnetic dips are not necessary and the filament threads are just observational manifestations of dynamical counter-streaming flows \citep{2001ApJ...553L..85K, 2016ApJ...831..123Z, 2017ApJ...836..122Z,2021ApJ...920..131G}.

At present, how the FC is formed is still a hot debated question. Many models have been proposed to interpret the formation of FC including the flux emergence model \citep{2008ApJ...673L.215O, 2014LRSP...11....3C, 2019ApJ...871...67C}, flux cancellation model \citep{1985AuJPh..38..929M, 1989ApJ...343..971V, 2000ApJ...539..983V, 2001ApJ...558..888G, 2015ApJ...809...83K, 2016ApJ...816...41Y}, and helicity condensation model \citep{2006ApJ...640..335A, 2010A&A...516A..49T, 2013ApJ...772...72A}. The emergence model assumes a highly twisted flux rope existing in the convective zone, which partially emerges to the corona via magnetic buoyancy. The emerging flux then forms an MFR through magnetic reconnection driven by shearing and converging flows \citep{2017ApJ...850...95S, 2017ApJ...850...39T}. In the flux cancellation model, the initial configuration is supposed to be a potential field. As the shearing motion along the PIL and converging flows toward the PIL are introduced, the potential field lines are first sheared and then reconnected to form two groups of new fluxes, one of which becomes longer and more twisted and the other is shorter and close to the potential. Afterwards, the short flux submerges to below the photosphere, appearing as flux cancellation \citep{1989ApJ...343..971V}. Obviously, in the two models, the sheared field and magnetic reconnection are two common ingredients, while the only difference is the origin of the sheared flux. However, the helicity condensation model is completely distinct from above the two models. It is essentially an accumulation of magnetic shear through the inverse helicity cascading, during which helicity is injected into the coronal flux by photospheric motions and flux emergence and submergence \citep{2010A&A...516A..49T, 2013ApJ...772...72A}.

The flux cancellation is found to be prevalent during the formation of the pre-eruptive MFR (e.g., \citealp{2007ApJ...666.1284W, 2008SoPh..248...51M, 2019ApJ...871...67C}). \citet{2008ApJ...673L.215O, 2009ApJ...697..913O} studied a sequence of vector magnetograms and found that the region with adjacent opposite polarities, where the horizontal magnetic field is strong but the vertical field is weak, first widens and then becomes narrow. Meanwhile, the reversal of the direction of the horizontal field along the PIL, the blueshift of spectral lines, and diverging flows are observed. The authors argued that these features show strong evidence for the MFR emergence model. However, such an interpretation was subsequently challenged by \citet{2012SoPh..278...33V}. They pointed out that the flux cancellation model is also able to give rise to the same observational features. Recently, it was disclosed that the small-scale flux cancellation is even more common than we previously realized, and that it may even drive nanoflares to heat the chromosphere and corona \citep{2018A&A...615L...9C, 2018MNRAS.479.2382L, 2018ApJ...862L..24P}.   

In this paper we study the growth of an active region (AR) NOAA 12790 FC with H$\alpha$ images of high spatial and temporal resolution. The interesting result is that the FC is gradually built up by a number of small-scale reconnection events occurring intermittently above the flux cancellation site prior to the eruption, at least during the time period we studied. This is also an important supplement for previous argument that the pre-eruptive MFR can be quickly formed by preceding confined major flares \citep{2013ApJ...764..125P,2018ApJ...867L...5L}. Section \ref{observations} describes instruments. The main results are presented in Section \ref{results}, which is followed by a summary and discussions in Section \ref{conclusion}.       

\section{Instruments}\label{observations}
The FC we study was primarily observed by the 1 m New Vacuum Solar Telescope (NVST; \citealp{2014RAA....14..705L, 2020ScChE..63.1656Y}), which is located at the Fuxian Lake of Yunnan Province, and operated by Yunnan Observatories. The NVST is designed to observe the fine structures in the lower solar atmosphere and reveal the origin and mechanism of the solar activities with high spatial ($\sim$0.165$^{\prime\prime}$ per pixel) and temporal resolution ($\sim$12 s). Due to limited field of view (FOV; $\sim$3$^{\prime}$), the NVST only observes a part of the FC. Fortunately, the Optical and Near-Infrared Solar Eruption Tracer (ONSET; \citealp{2012EAS....55..349F, 2013RAA....13.1509F, 2015ApJ...809...46C}), also located at the Fuxian Lake, has a larger FOV and thus enables us to observe the entire FC. Moreover, we also use the data from the Atmospheric Imaging Assembly (AIA; \citealp{2012SoPh..275...17L}) onboard Solar Dynamics Observatory (SDO; \citealp{2012SoPh..275....3P}), which provides the full solar disk images in 7 extreme ultraviolet (EUV) and 2 ultraviolet (UV) passbands. The temporal cadence is 12 s (24 s) for the EUV (UV) passbands. The pixel size is 0.6$^{\prime\prime}$ for both. The Helioseismic and Magnetic Imager (HMI; \citealp{2012SoPh..275..229S, 2012SoPh..275..207S}) onboard SDO provides line-of-sight (temporal resolution $\sim$45 s) and vector magnetograms (temporal resolution $\sim$12 min), which are utilized to investigate the temporal variation in magnetic flux and reconstruct the three-dimensional (3D) coronal magnetic field, respectively.
 
\section{Results}\label{results}
\subsection{Grow-up of Filament Channel and Mass Transfer}
\begin{figure*}
	\includegraphics[width=16cm,clip]{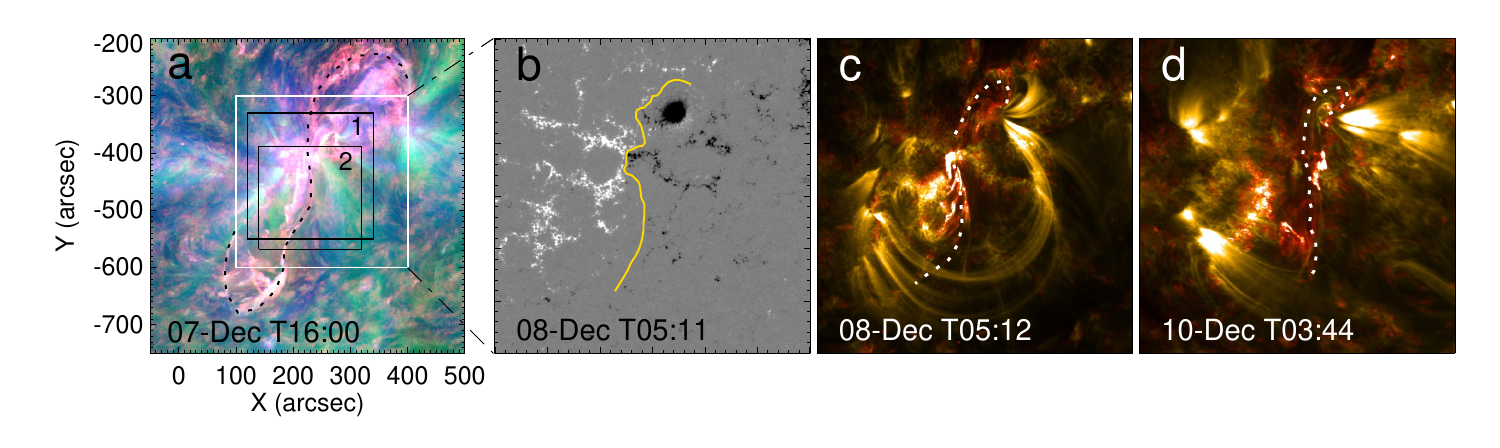}
	\centering
	\caption{a: Composite of AIA 304 \AA, 211 \AA, 171 \AA\ image showing the filament eruption on 2020 Dec. 7. The white box shows the FOV of panel b-d. The black boxes 1 and 2 show the FOVs of the ONSET and NVST images as shown in Figure \ref{figure2}b and Figure \ref{figure3}, respectively. b: HMI line-of-sight magnetogram with the main PIL indicated by the gold line. c-d: Composite of the AIA 304 \AA\ and 171 \AA\ images showing the remained FC and the confined filament eruption. The dashed line indicate the corresponding FC at different instants.  \label{figure1} }
\end{figure*}

On 2020 Dec. 07, the filament originating from AR 12790 produced the first halo CME in the 25th solar cycle, as shown in Figure \ref{figure1}a (also see \citealp{2021ApJ...919...34Y}). Prior to the eruption, the filament threads were highly sheared and the corresponding magnetic configuration, defined as the FC, was expected to be mainly aligned with the main PIL of photospheric magnetic field (the gold line in Figure \ref{figure1}b), which was composed of a leading negative polarity and some dispersed positive polarities. After the major eruption, partial FC remained at the south of the PIL, and manifested as sheared filament threads (Figure \ref{figure1}c). During the following time period of more than 40 hours, the FC seemed to continuously grow and finally appeared as a well observed filament on 2020 Dec. 10. At approximately 03:50 UT, the FC erupted again but did not succeed (Figure \ref{figure1}d), only causing an A-class flare.

\begin{figure*}[htbp]
	\includegraphics[width=15cm,clip]{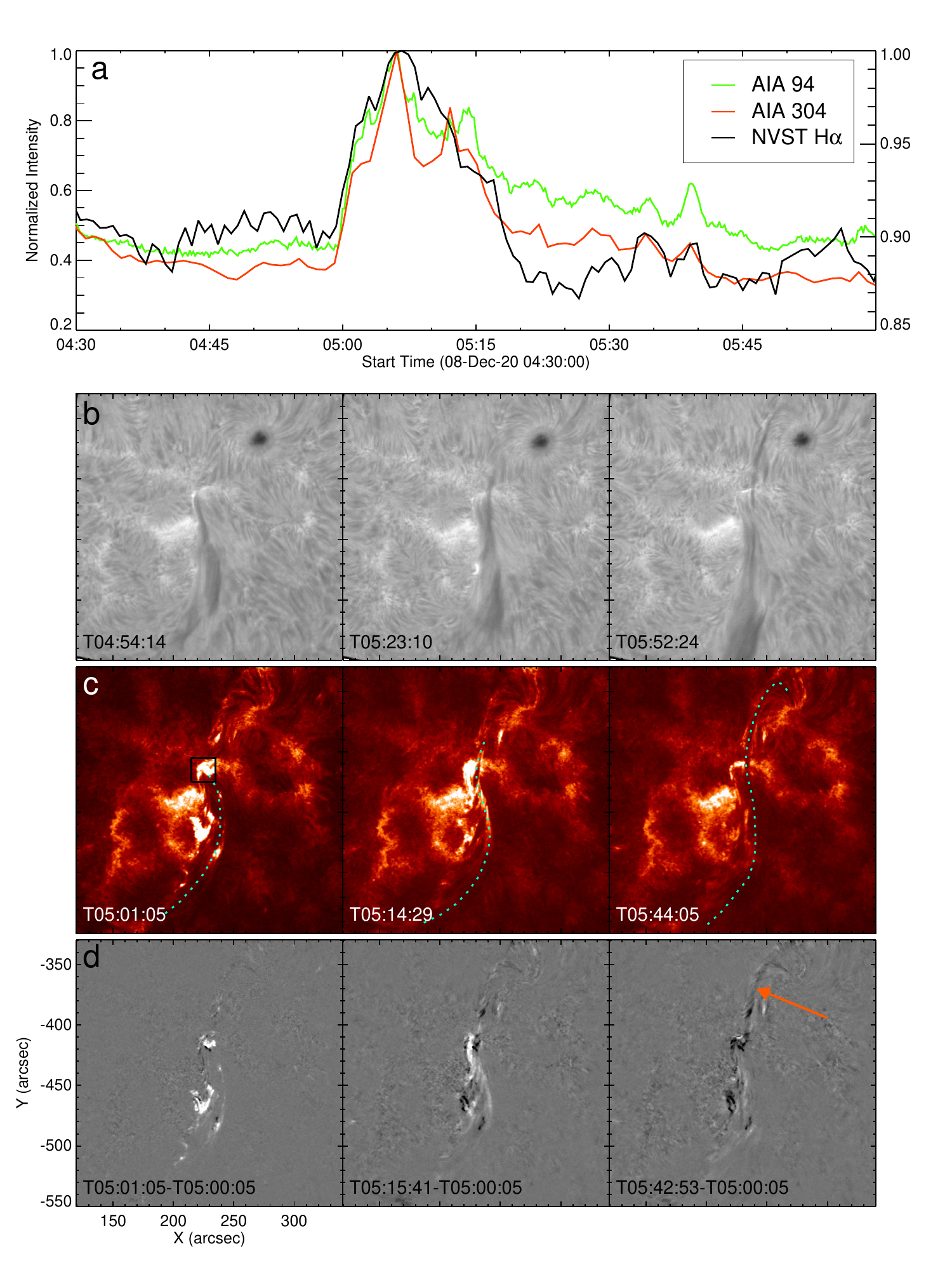}
	\centering
	\caption{a: Temporal evolution of the normalized AIA 94 \AA, 304 \AA\ and H$\alpha$ intensities within the black box in panel c. b: ONSET H$\alpha$ images showing the formation of the filament threads. They are aligned with the AIA images through cross-correlation between the ONSET and HMI white-light images. c and d: AIA 304 \AA\ and corresponding running-difference images. The dashed line represents the filament threads, and the orange arrow points out the transfered filament. \label{figure2}}
\end{figure*}

In this paper, we mainly focus on one episode of the FC build-up after the first major eruption, because it happens to be simultaneously observed by the NVST and ONSET. At 05:00 UT on 2020 Dec. 08, an H$\alpha$ burst appeared as a small-scale brightening near the filament north end. Afterward, some new filament threads were formed with their northern ends extending to the penumbra of the preceding sunspot at 05:52 UT (Figure \ref{figure2}b). The small-scale H$\alpha$ burst, as well as the accompanied dynamics of the filament threads, also caused a response at the AIA EUV passbands (Figure \ref{figure2}a). As shown in Figure \ref{figure2}c, with the commence of the EUV brightening, one can see that partial filament materials (the dotted line in Figure \ref{figure2}c) were transferred from south to north of the main PIL. At 05:44 UT, some continuous and long filament threads obviously appeared at the AIA 304 \AA\ passband, highly resembling what was observed in H$\alpha$. It was also apparently observed from the running-difference images as shown in Figure \ref{figure2}d. During the same time period, some heated plasma blobs were also observed to be quickly ejected from the brightenting site along two opposite directions parallel to the PIL, suggestive of the occurrence of magnetic reconnection.

\begin{figure*}[htbp]
	\includegraphics[width=15cm,clip]{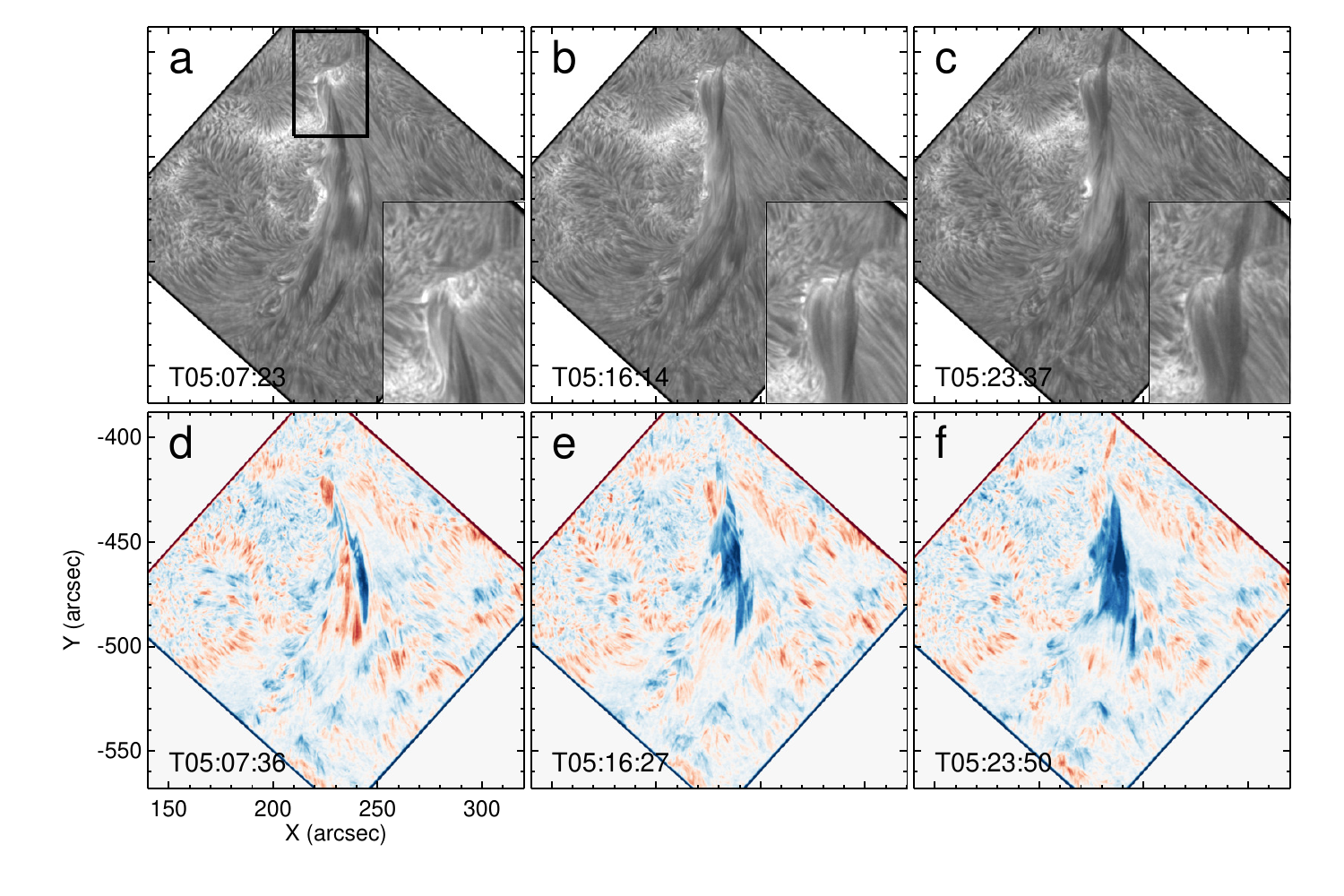}
	\centering
	\caption{a-c: NVST H$\alpha$ images showing the FC evolution and grown to the northern part. The zoom in of the region of interest is shown in the lower right corner of each panel (An animation is available online). d-f: Corresponding pseudo Doppler maps.  \label{figure3}} 
\end{figure*}

Regardless of limited FOV of the NVST, it exactly covered the H$\alpha$/EUV burst. From the H$\alpha$ images of high spatio-temporal resolution (Figure \ref{figure3}a-\ref{figure3}c) and the attached movie, one can clearly observe the transfer of the filament materials from arc-shaped arcades to the north part of the PIL after the brightening occurred. Before and during the transfer process, the southern arc-shaped filament was even found to present an untwisting motion. The rotation direction was clockwise when observed from the south of the AR, indicating that the magnetic structure of the southern arc-filament reserved a certain amount of twist, at least being highly sheared, as suggested by \citet{2015NatCo...6.7008W}. The corresponding chirality is sinistral, or the helicity is positive, which is consistent with the hemispheric rule of magnetic helicity \citep{1992ASPC...27...53M, 1998SoPh..182..107M, 2014ApJ...784...50C, 2017ApJ...835...94O}. We conjectured that the reconnection most likely took place between the arc-shaped filament system and the nearby flux above the northern PIL. It formed longer magnetic field lines, almost spanning the entire PIL, and gave rise to the H$\alpha$/EUV burst at the same time. As the long field lines were filled with transferred filament materials, they then became visible as dark threads.

The H$\alpha$ off-band observations of the NVST provided Doppler shift maps during the formation of the FC. For this particular event, the wavelengths of H$\alpha$ off-band were centered at $6562.8 \pm 0.4$ \AA. The formula we used to calculate Doppler maps was from \citet{2008ApJ...673.1201L}, as shown below :
\begin{equation}
	D=\frac{B-R}{B+R}
\end{equation}
where $B$ and $R$ represent the blue-wing and red-wing intensities, respectively. The results were shown in Figure \ref{figure3}d-\ref{figure3}f, from which one can see that the counter-streaming along the arc-shaped FC instantly became visible once the H$\alpha$/EUV brightenings appeared. However, shortly afterwards, the blue-shift started to dominate (Figure \ref{figure3}e-\ref{figure3}f), which was a result of the combination of the untwisting motion of the filament magnetic structures and the field-aligned motion of the materials from the leg to top of the field lines. The latter was most likely driven by the reconnection outflows. 
 
\begin{figure*}[htbp]
	\includegraphics[width=15cm,clip]{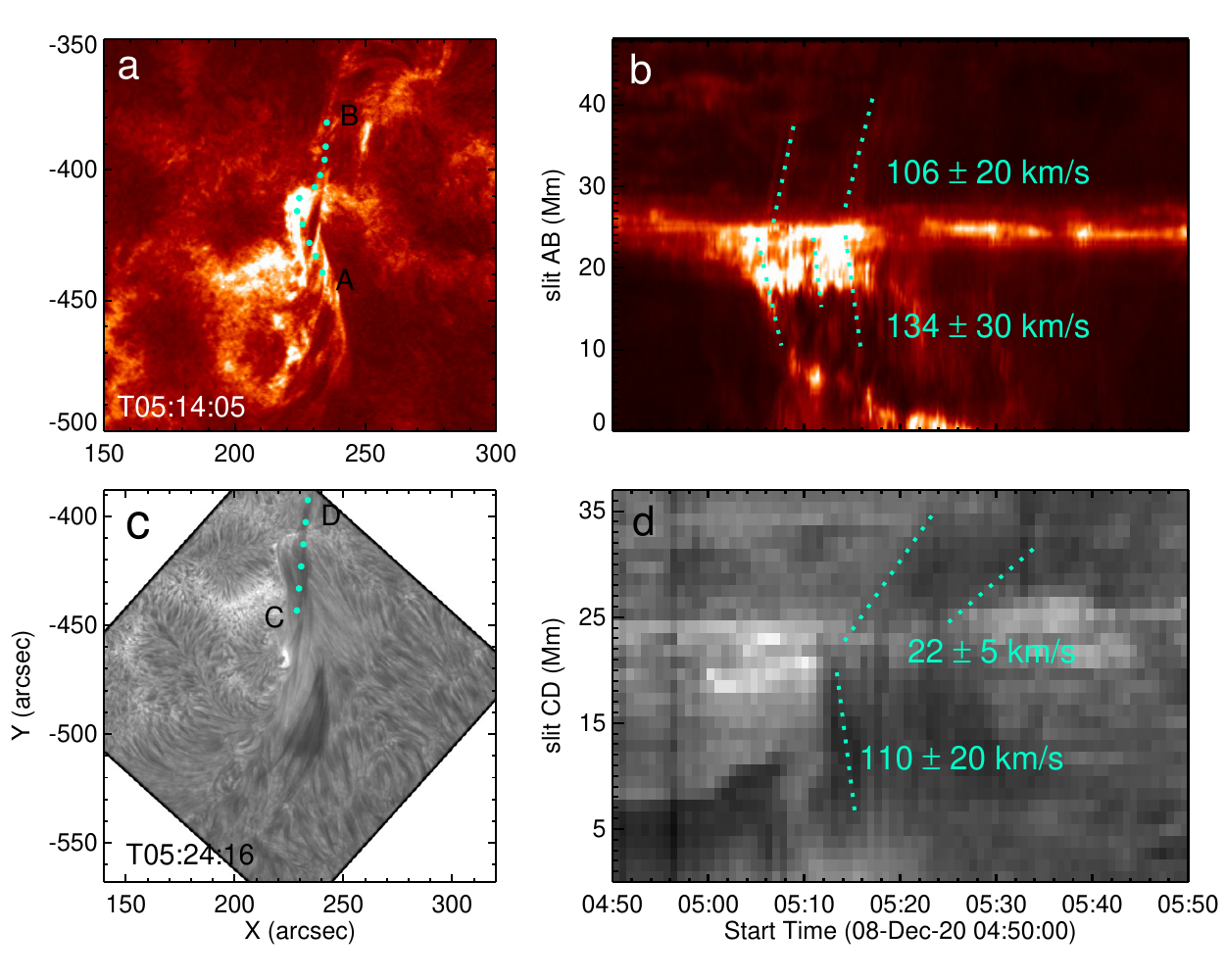}
	\centering
	\caption{a and c: AIA 304 \AA\ and NVST H$\alpha$ images. The curved slits AB and CD, as shown by two dotted lines, represent the directions of the reconnection outflows. b and d: 304 \AA\ and H$\alpha$ slice-time plots. The inclined dotted lines show the trajectories of outflow jets. \label{figure4}} 
\end{figure*}

To quantify the motion of the bidirected outflows ejected by magnetic reconnection in the plane of sky, we took two curved slices AB and CD along the FC as shown in Figure \ref{figure4}a and \ref{figure4}c. The slice-time plots for the AIA 304 \AA\ and NVST H$\alpha$ passbands are displayed in Figure \ref{figure4}b and \ref{figure4}d, respectively. One can clearly see two groups of outflow jets that were launched after the appearance of the brightenings and then quickly moved toward the opposite directions. At the AIA passbands, the outflow jets were only visible at the 304 \AA\ passband, inferring that they were not heated to the coronal temperatures (e.g., above 1 MK). Through tracking the trajectories of the outflows, the velocities were estimated to be in the range of $100-150$ km s$^{-1}$. Similar to the AIA 304 \AA\ passband, at the H$\alpha$ passband, one can observe some quickly moved dark jets, while the corresponding velocities were slightly smaller. 

\subsection{Causes and Intermittency of Magnetic Reconnection}
\begin{figure*}[htbp]
	\includegraphics[width=15cm,clip]{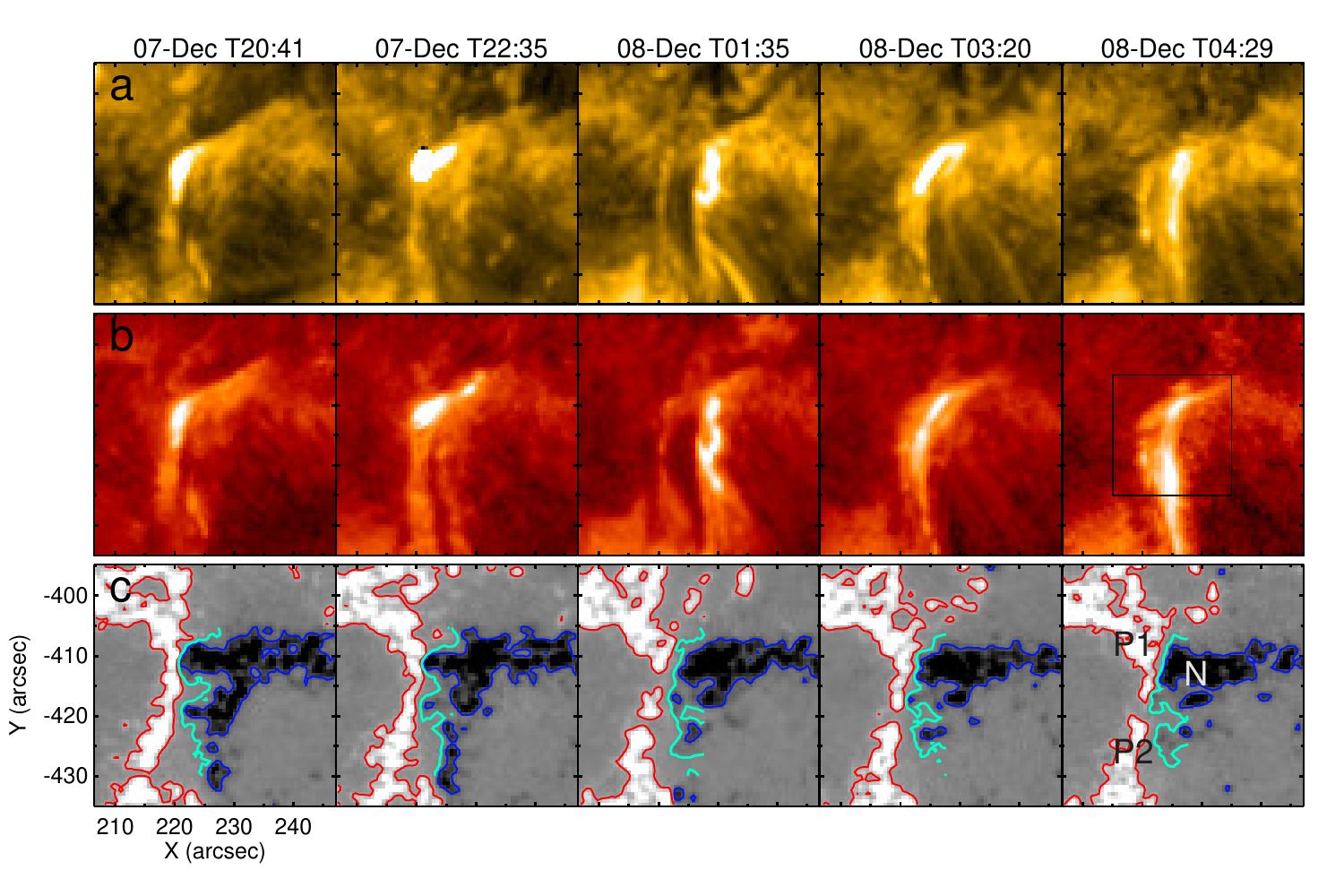}
	\centering
	\caption{a-b: Time-sequence of AIA 171 \AA\ and 304 \AA\ images displaying repeatedly occurred EUV bursts at the site of the flux cancellation. The black box shows the region for integrating the intensity. c: HMI LOS magnetograms with red/blue contours representing the positive/negative magnetic fields of ±100G. The cyan line represents the PIL, and P1 (P2) and N represent the positive and negative polarities, respectively.  \label{figure5}}
\end{figure*}

\begin{figure*}[htbp]
  \includegraphics[width=15cm,clip]{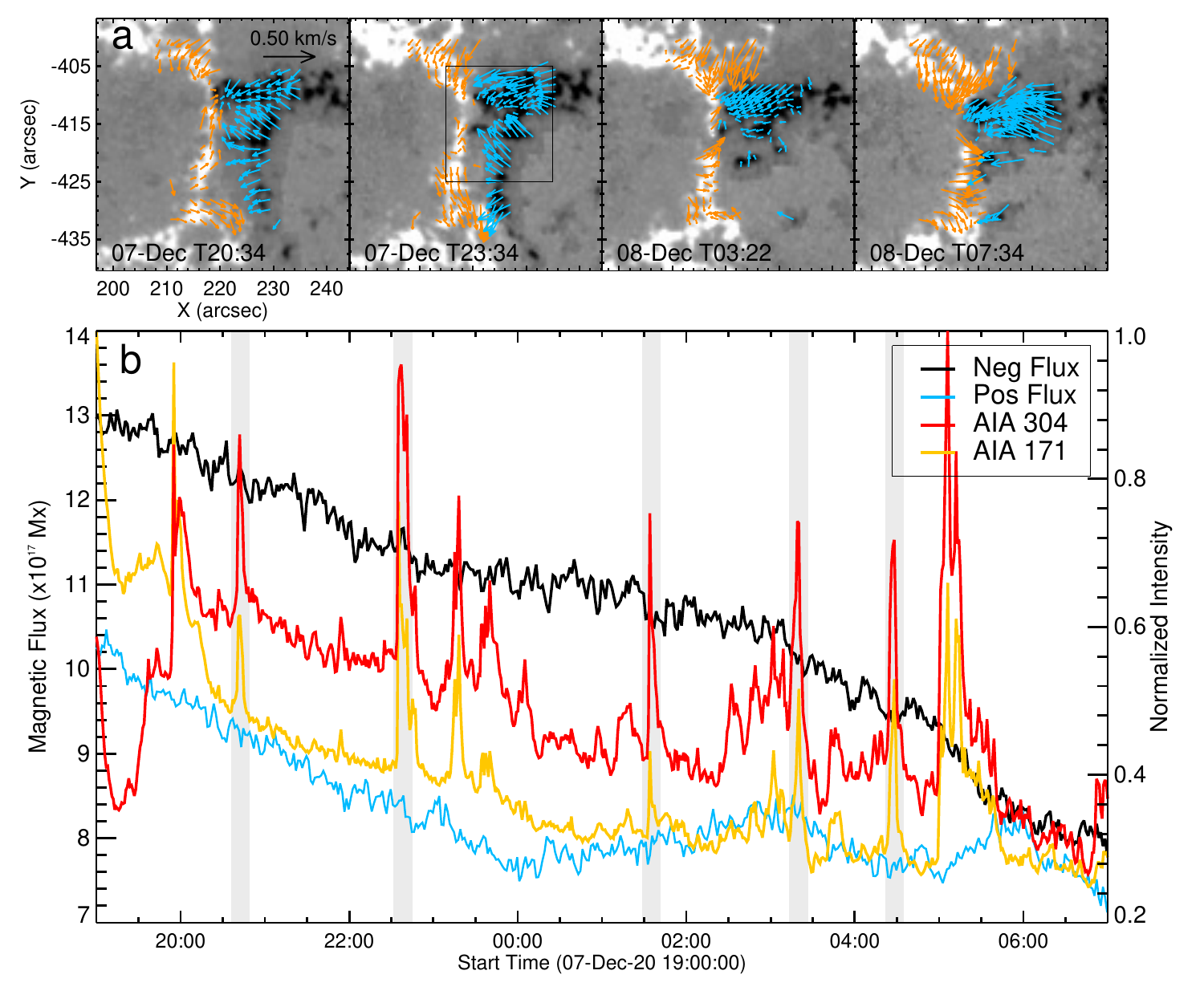}
  \centering
\caption{a: HMI LOS magnetograms superimposed with the velocity fields. The blue (orange) arrows represent the velocity of the negative (positive) polarities. b: Temporal evolution of integrated magnetic fluxes and normalized 304 \AA\ and 171 \AA\ intensities. The black box shows the region for integrating magnetic fluxes. The shadowed regions correspond to the time periods of bursts shown in Figure \ref{figure5}. \label{figure6}}
\end{figure*}

We speculated that the formation of the FC prior to the eruption may experience multiple reconnection episodes. That is to say, a single reconnection event, as revealed by the H$\alpha$/EUV burst, may not be able to supply enough magnetic flux to trigger the eruption (e.g., \citealp{2014ApJ...789..133Z, 2020Innov...100059X}). Through inspecting the long-term evolution of the AIA 304 \AA\ and 171 \AA\ images, we found 9 EUV bursts, at least, appearing at the same location repeatedly and presenting a similar morphology, as shown in Figure \ref{figure5}a and \ref{figure5}b. Each of them lasted for about 10 minutes, even though their magnitudes changed from case to case, as shown by the evolutions of the 304 \AA\ and 171 \AA\ intensities (Figure \ref{figure6}b). This implies that the formation of a full-fledged FC needs to experience multiple reconnection processes, which take place intermittently and manifest as the intermittent appearance of EUV bursts. 

The H$\alpha$/EUV burst was found to be driven by the shearing and converging flows at the photosphere. Figure \ref{figure5}c shows a time-sequence of the MHI line-of-sight magnetograms. We found that the EUV bursts were most likely related to the interaction of two parasitic positive and negative polarities. To confirm such an interaction, we inspected the distribution and evolution of the velocity field in the photosphere, which was estimated by the DAVE4-VM method proposed by \citet{2008ApJ...683.1134S} using the HMI SHARP vector data. It was clear that, in such a small-scale region, the negative polarities (N) were continuously moving toward to the positive ones (P1, P2) with a maximum velocity of 0.5 km s$^{-1}$ (Figure \ref{figure6}a). At the location of the H$\alpha$/EUV burst, they ceaselessly cancelled with each other and gave rise to Ha/EUV bursts. The flux cancellation was also revealed by the integrated negative flux, which showed a gradual decrease with time going on, as shown by the black curve in Figure \ref{figure6}b. The total cancelled negative flux during 12 hours was estimated to be about $5 \times 10^{17}$ Mx, with an average cancellation rate of $\sim$$10^{13}$ Mx s$^{-1}$. Comparing with the flux cancellation rate derived by \cite{2021A&A...647A.188D}, the rate derived here seems to be not enough for a strong reconnection event. At the same time, the positive flux also tended to decrease but had an increases during two intervals (the blue curve in Figure \ref{figure6}b). The increased flux was mainly caused by that the positive polarity out of the region where we integrated the flux continuously flowed into.

\subsection{3D Magnetic Property of  Filament Channel Build-up}
\begin{figure*}[htbp]
	\includegraphics[width=15cm,clip]{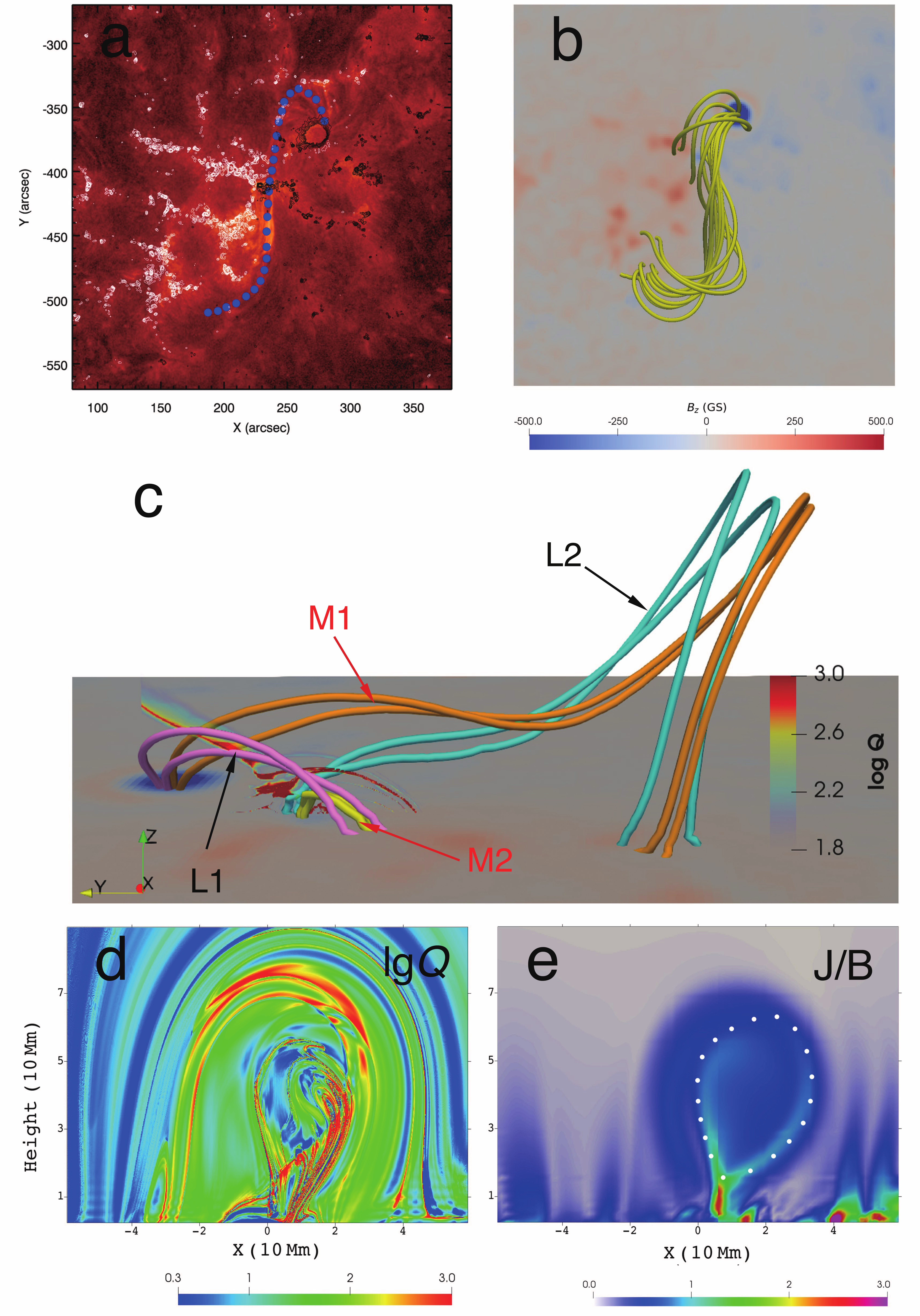}
	\centering
	\caption{a: AIA 304 \AA\ image with the dotted curve overlaid showing the path of the FC. b: The collection of 3D magnetic field lines indicating the MFR as seen from the top. The bottom boundary is HMI LOS magnetogram. c: The side view of the panel b coupled with $Q$ values. L1 and L2 delineate two groups of sheared arcades, and M1 and M2 represent the MFR field lines and small flare loops, respectively. The coordinates x/y/z represent the west/north/altitude. d: The distribution of $Q$ values on the plane perpendicular to the MFR axis, as pointed out by the dashed line in the panel b. e: The distribution of $J/B$ in the same plane and the dotted line indicates the outer boundary of the MFR.  \label{figure7}}
\end{figure*}

To understand 3D magnetic property of the FC, we reconstructed the coronal magnetic field of the source region based on the non-linear force-free field (NLFFF) assumption using the HMI photospheric vector magnetogram as the bottom boundary \citep{2001ApJ...551L.115Y, 2009ApJ...693L..27C, 2012LRSP....9....5W, 2016ApJ...824..148Y, 2019ApJ...871..105Z, 2020ApJ...901...13Q}. We first tried the optimization extrapolation method \citep{2010ApJ...714..343G, 2016ApJ...826...51Z} and found that it was difficult to reproduce the magnetic field comparable with the morphology of the observed filament threads. We then resorted to the magnetic flux rope (MFR) embedding method based on regularized Biot-Savart laws (RBSL; \citealt{2014ApJ...790..163T, 2018ApJ...852L..21T,  2019ApJ...884L...1G}). The procedure was divided into four steps. First, we took advantage of the 304 \AA\ images at 05:24 UT to derive the path of the FC, as shown in Figure \ref{figure7}a. According to previous statistics \citep{Tanberg1995,Filippov2000,Engvold2015}, the height of active region filaments ranges from 5 Mm to 30 Mm. We thus set the height of the MFR axis to be 30 Mm and the MFR minor radius to be 25 Mm assuming that the lower half part of the MFR are fully filled by cool materials. The 3D path of the MFR was estimated according to \citet{2021ApJ...917...81G}. Second, we calculated a potential field utilizing the radial magnetic field component, where the projection effect was corrected \citep{2006SoPh..233..215W, 2017ScChD..60.1408G}. Third, we set the physical parameters of the RBSL model including the average value of unsigned magnetic flux at the two MFR footprints ($1.72 \times 10^{21}$ Mx) and the strength of the electric current following the equation (12) in \citet{2018ApJ...852L..21T}. Finally, we inserted the MFR derived by the RBSL model into the potential field along the path of the FC and then performed a relaxation using the magnetofrictional code \citep{ 2016ApJ...828...83G, 2016ApJ...828...82G}. After relaxation, the force-free metric was $\sigma_{J}=0.28$, and the divergence-free metric was $\langle |f_{i}|\rangle =1.47 \times 10^{-4}$, which were small enough and basically acceptable according to \citet{2021ApJ...917...81G}\footnote{\href{URL}{https://github.com/njuguoyang/magnetic\_modeling\_codes}.}.

Figure \ref{figure7}b shows selected magnetic field lines of the AR NLFFF structure, from which one can see that there existed two groups of highly sheared field lines underneath the modeled MFR (M1) as indicated by L1 and L2 in Figure \ref{figure7}c. The right leg of L1 and the left leg of L2 formed an X-shaped configuration. Their footpoints were both rooted in the region where the converging motion and flux cancellation took place. At the same time, we also observed some much shorter loops that were located below the X-shaped configuration (M2 in Figure \ref{figure7}c). Such a configuration was consistent with the tether-cutting reconnection model \citep{1980IAUS...91..207M, 2001ApJ...552..833M}, in which the reconnection of two sheared arcades formed a longer and twisted loop above and a shorter semicircular-like loop below. Owing to the magnetic tension, the flux M1 would rise up. Because the southern part of M1 was highly inclined, the reconnection outflows toward the south would also produce an upward velocity, appearing as the blueshifts in the NVST off-band images. Meanwhile, the northern part was low-lying, the outflows toward the north appeared as redshifts at the location near [240, -400]. On the other hand, due to the downward magnetic tension, M2 would submerge, and thus manifested as the flux cancellation.

To further quantify the property of the reconnection, we calculated the squashing factor ($Q$) and the current density on the plane almost perpendicular to the inserted MFR axis and cutting through the reconnection X-point as shown in Figure \ref{figure7}d and \ref{figure7}e. The isosurfaces of high $Q$ values indicated quasi-separatrix layers (QSLs), which describe the locations of rapid magnetic connectivity changes \citep{1995JGR...10023443P, 2002JGRA..107.1164T}. We found that the MFR was well wrapped by the QSLs and that an obvious hyperbolic flux tube (HFT) was formed below the MFR. The current density was also found to be the strongest at the HFT. These features thus strengthened our previous conjecture for the occurrence of magnetic reconnection. However, it is worth of mentioning that the height of the reconnection X-point (about 12 Mm) in the NLFFF structure has a large uncertainty. First, it may be caused by the NLFFF assumption, which did not consider the chromosphere and transition region that may include strong Lorenz force. Second, the parameters of the inserted MFR were set by experience. When we changed the height and radius of the MFR by $\pm{5}$ Mm, the height of reconnection site varied from 5 to 18 Mm. Fortunately, the AR was also observed by the IRIS, the spectroscopic data of which allowed us to further explore the properties, in particular the height, of the reconnection responsible for the FC formation. The results will be present in a separated paper. Moreover, it was found that the derived MFR transited to an SMA during the relaxation if the average twist of the inserted MFR was turned down. However, the reconnection between two sheared arcades was still observed.

To strengthen our argumentation that the FC was gradually built up by the mutilple reconnection events that occurred intermittently, we compared the two NLFFF models before and after the intermittent reconnection events we observed. For the two models, all initial input values are the same except for the bottom magnetic field. To uncover the temporal variation of the FC, we needed to calculate the twist and toroidal flux of the MFR. We first identified the boundary of the MFR (Figure \ref{figure7}e) by using the IDL routine \textit{region\_grow.pro}\footnote{\href{url}{https://www.l3harrisgeospatial.com/docs/REGION\_GROW.html}.}. This method searches for an MFR region starting from a selected small region near the MFR center and determining which neighbor pixels should be added to the region. We then adjusted the location of the small region and repeated the same procedure by 7 times. Finally, using the code of \citet{2016ApJ...818..148L}\footnote{\href{URL}{http://staff.ustc.edu.cn/~rliu/qfactor.html}.}, we calculated the twist and toroidal flux of magnetic field surrounded by the identified boundary and took their averages as eventual values. The uncertainties were the corresponding standard deviations, as listed in Table \ref{table1}. It was found that the two values both increased during the FC grow-up as we expected. Furthermore, we noticed that the average twist of the MFR is smaller than one, indicating the low twisted magnetic lines dominated the MFR. This might be the primary reason for the FC eruption being confined on 2020 Dec. 10.

\begin{table*}[htbp]
 \centering
 \setlength{\tabcolsep}{10mm}
 \renewcommand{\arraystretch}{2}
 \caption{Average twist numbers and toroidal fluxes for the NLFFF models at two moments. }
	\begin{tabular}{c|c|c}
	\hline 
	Time (UT) & Average Twist & Toroidal Flux (Mx) \\ 
	\hline
	07-Dec 20:36:00 & $0.86 \pm 0.01$ &  $1.09  \pm 0.12 \times 10^{21}$  \\  
	\hline
	08-Dec 05:12:00 & $0.88 \pm 0.03$ &  $1.52 \pm 0.24 \times 10^{21}$   \\
	\hline
\end{tabular}\label{table1}
\end{table*}

\section{Summary and Discussion} \label{conclusion}
In this paper, we studied the growth of a FC in the AR NOAA 12790. We found that the FC was gradually built up by a series of small-scale reconnection events, which manifested as repeated H$\alpha$/EUV bursts. The H$\alpha$/EUV emission enhancement, bi-directional outflow jets, and untwisting motion during the observed burst provided strong evidences of magnetic reconnection. Thanks to the NVST data of high spatio-temporal resolution, It was observed that the filament materials were partly and quickly transferred to longer and more twisted magnetic field lines, which were most likely formed during the burst. The NLFFF of the AR using the MFR embedding method further disclosed the reconnection configuration, which was composed of two groups of SMAs and an HFT embedded in between. As the HFT reconnection occurred, the long and twisted flux was gradually accumulated, resembling the FC. This process also produced the short loops during the same period, which subsequently submerged appearing as the small-scale flux cancellation. The horizontal velocity field in the photosphere further disclosed the driver of the HFT reconnection, i.e., the continuous shearing and converging flows near the flux cancellation site.
 
The FC build-up disclosed here is in favor of the flux cancellation model proposed by \citet{1989ApJ...343..971V}. In addition, we uncovered the two new features that were not specified in this model. The first one is that the reconnection configuration is of the HFT type rather than of the bald-patch type as indicated in the model. The second and more interesting one is that the HFT configuration and thus the HFT reconnection are intermittent rather than continuing even though the driving flows and flux cancellation seem to be continuous. We thus suspect that the reconnection configuration was highly dynamic over time. During the non-burst periods, the reconnection perhaps was extremely weak and the corresponding configuration also changed into a bald-patch, which only allowed the magnetic dissipation in the photoshpere, giving rise to the flux cancellation. On the other hand, it was noticed that the seed flux building the FC we study was from the remains of the previous eruption on 2020 Dec. 7. This reminded us that, for a full-fledged FC, its formation may be more complicated, most likely involving multiple mechanisms.

The transfer of materials is an important indicator of the reorganization of the magnetic field through magnetic reconnection. Such a phenomenon has often been observed, in particular, during the eruption of filaments. \cite{2019ApJ...887..239Y}  found that, during the eruption of a mini-filament, an obvious untwisting of the erupting dark threads and a rotation motion of the associated blowout jets in the nearby large-scale loops appeared simultaneously. Therefore, they proposed that the reconnection took place between the filament flux and the background field and rapidly transferred magnetic twist from the former to the latter, as delineated in their Figure 8. Such a twist transfer process may be more common during the confined/failed filament eruption for the sake of re-distributing magnetic helicity (e.g., \citealp{2020ApJ...904...15Y, 2020ApJ...889..106Y}). The event studied here further revealed that the material transfer and twist transfer also occurred during the grow-up of the FC. For each H$\alpha$/EUV burst event, the flux injected to the FC may be very limited. However, with a number of such small-scale events during a long time period, the FC can be easily built up and ready for eruption.  

In fact, the build-up of the FC through the intermittent reconnection has been indicated by \cite{2013ApJ...779..157G}. Through investigating the evolution of magnetic helicity and twist of a flare-productive active region, they concluded that the confined flares prior to the eruptive one played an important role in building up the eruptive MFR. These confined flares that intermittently occurred near the PIL were similar to the H$\alpha$/EUV bursts observed here, while the only difference was that the magnitude of the latter was much smaller than that of the former. It seems that the larger the burst magnitude is, the more the flux added to the FC (e.g., \citealp{2018ApJ...867L...5L}). On the other hand, \cite{2013ApJ...779..157G} identified that the reconnection took place in the whole QSLs that demarcated the MFR from the ambient field. However, for the current case, the reconnection was demonstrated to occur in the more localized HFT between the MFR and the flux cancellation site. \\

\begin{acknowledgements}
We thank the referee for his/her careful reading to improve the clarity of the manuscript. We also thank the NVST, ONSET, SDO/AIA, and SDO/HMI teams for providing the high-cadence data. This work is supported by NSFC grants 11722325, 11733003, 11790303, 11873087, Yunnan Science Foundation for Distinguished Young Scholars No. 202001AV070004 and National Key R\&D Program of China under grant 2021YFA1600504. X.C. is also supported by Alexander von Humboldt foundation.
\end{acknowledgements}

\bibliographystyle{aa}
\bibliography{reference}

\end{document}